\documentclass[preprint2]{aastex701}
\begin{document}

\title{Evaluating Cepheid Metallicity Effect Determinations via IC1613 and Gaia-independent Parallaxes}

\author[orcid=0000-0001-8803-3840]{Daniel Majaess}
\affiliation{Mount Saint Vincent University, Halifax, Nova Scotia, Canada.}
\email[show]{Daniel.Majaess@msvu.ca}

\begin{abstract}
18 HST parallaxes for Galactic classical Cepheids are unified to establish a Gaia-independent IC1613 distance and evaluate metallicity effect determinations, since the Gaia zeropoint is debated. Recently proposed classical Cepheid metallicity corrections of $\gamma (W_{VI}) \simeq 0, -0.25, -0.50$ mag dex$^{-1}$ are benchmarked, and yield $\mu_{0,{W_{VI}}} \simeq 24.39, 24.16, 23.93$ ($\pm0.07$).  Larger corrections are disfavored relative to a weighted mean of IC1613 TRGB and TRGB/JAGB distances of $24.39^{+0.07}_{-0.04}$ (EDD) and $24.36\pm0.06$/$24.45\pm0.11$ (CCHP). A more expansive metallicity baseline is desirable to scrutinize smaller corrections (e.g., $\gamma (W_{VI}) \lesssim -0.1$ mag dex$^{-1}$), while concurrently acquiring additional non-Gaia parallaxes since published concerns exist regarding DR3 (e.g., critical long-period Cepheids S Vul, SV Vul).
\end{abstract}


\section{Introduction}
 The present effort was undertaken to benchmark larger metallicity $\gamma (W_{VI})$ corrections to classical Cepheid distances, since recently published determinations span a $\Delta \gamma (W_{VI}) \simeq 0.5$ mag dex$^{-1}$ baseline.  IC1613 is leveraged to constrain $\gamma (W_{VI})$ because it is metal-poor relative to the Milky Way \citep[e.g.,][and references therein]{tal81}. This research likewise aims to bypass brighter Gaia DR3 parallaxes since there is an ongoing debate concerning a floating zeropoint \citep[e.g., Fig.~10 in][Figs.~6, 7, and 8 in \citealt{kha23}, and see also the contrasting positions between \citealt{owe22} and \citealt{bre22}]{mol23}.  A $W_{VI}$ distance to IC1613 is instead determined via 18 Gaia-independent HST parallaxes from \citet{rie14,rie18plx} and \citet{ben02,ben07}, thereby providing advantageous independent insight.  Finally, it was recently implied that a broad consensus was converging toward how composition impacts the Cepheid distance scale, whereas alternative narratives emerge here and by \citet{rip25a}.  Indeed, diverse interpretations exist from comparatively null to sizable adjustments \citep[e.g.,][]{pie04,sha11,mf23,rip25}.

\section{Analysis}
Photometry for IC1613 and Milky Way Cepheids stemmed from \citet{uda01} and \citet[][]{nge12}\footnote{See also \citet{ber00}.}, and the Galactic parallaxes are tied to HST \citep{ben07,rie14,rie18plx}.  The ratio of total to selective extinction utilized was employed by \citet{bre22} ($R_{V-VI}=A_V/E(V-I)\simeq2.39$). Numerous complexities encompass the selection of $R$, and the topic is discussed elsewhere \citep[][see also Table 6 in \citealt{rie22} and the Appendix in \citealt{mad25}]{tu76,uda03,wh24,maj25e}. Cepheids $r< 2$' from the core and featuring $P<5^{d}$ were cropped owing to contamination concerns \citep{maj13,maj14}, in tandem with generally deviant Cepheids V39 and [UWP2001] 13709.  By design those former choices systematically increase the distance. Imposing a metallicity-independent\footnote{Figs.~2 and 1 in \citet{maj10} and \citet{maj11}, accordingly. However, Galactic and SMC Cepheids may adhere to differing BV Wesenheit slopes \citep{maj08,maj09c}, likely a result of line blanketing \citep[][and references therein]{cc85}.  See also the $W_{BV}$ interpretations of \citet{bm99}.} slope of $-3.30$ \citep{uda00} yields $\mu_{0,{W_{VI}}} = 24.39\pm0.07$ (no abundance correction).  The metallicity effect recently derived by \citet[][]{mf23,mf25}, \citet[][$-0.264\pm0.058$]{bre24}, and \citet[][$-0.52\pm0.19$]{rip25} are assessed by evaluating $\gamma (W_{VI}) \simeq 0, -0.25, -0.50$ mag dex$^{-1}$, which yield $\mu_{0,{W_{VI}}} = 24.39, 24.16, 24.93$ ($\pm0.07$) (Table~\ref{table1}). That can be compared to the TRGB estimate from \citet[][EDD, $24.39^{+0.07}_{-0.04}$]{ana21}, and TRGB-JAGB distances from \citet[][CCHP]{lee24} of $24.36\pm0.06$ and $24.45\pm0.11$.  The weighted mean disfavors larger corrections.  Additional literature estimates for IC1613 distances can be found in \citet[][their Table~6]{lee24}, \citet{mf25}, and \citet{bre25}.

\begin{table}[t]
\caption{Benchmarking recent Cepheid metallicity corrections via IC1613 distances:~$24.39^{+0.07}_{-0.04}$ \citep[TRGB,][EDD]{ana21}, $24.36\pm0.06$ and $24.45\pm0.11$ \citep[TRGB-JAGB,][CCHP]{lee24}.}
\label{table1}
\begin{center}
\begin{tabular}{cc}
\hline
$\gamma (W_{VI})$ & $\mu_{0,{W_{VI}}}$ IC1613 \\ 
 mag dex$^{-1}$ &  ($\pm0.07$) \\
\hline
$0$  & $24.39$ \\                    
$-0.25$ & $24.16$ \\
$-0.50$ & $23.93$ \\
$-0.75$ & $23.69$ \\
\hline
$-0.10$  & $24.30$ \\
\hline
\end{tabular}    
\end{center}
\textit{Notes}:~distances stem from a $W_{VI}$ relation tied to 18 HST parallaxes \citep{ben07,rie14,rie18plx}, since consensus is lacking regarding Gaia corrections, particularly for brighter targets \citep[e.g., Fig.~10 in][Figs.~6, 7, and 8 in \citealt{kha23}, and see also the contrasting positions between \citealt{owe22} and \citealt{bre22}]{mol23}. The IC1613 Cepheids utilized were selected from \citet[][OGLE]{uda01} following criteria in the text.
\end{table}

A $\rm{\Delta [Fe/H]}\simeq 0.93$ metallicity baseline is assumed and stems from comparing IC1613 B and A-type supergiants beyond the core \citep[][${\rm [Fe/H]}\simeq-0.85$]{ber18}, and a median for the Galactic sample based on the \citet[][]{gro18} compilation ($\rm{[Fe/H]}\simeq0.08$, and references therein). For cases where a third Galactic $\rm{[Fe/H]}$ datum was lacking for individual stars an additional literature estimate was utilized \citep{luc11,use13a,use13}.  B and A-type supergiant results are preferred relative to red supergiants given the dependence on multiple irons lines, and less reliance on $\alpha$-elements. \citet{dav15} employed red supergiants and yielded systematically higher metallicities for the SMC ($[Z]=-0.53$) relative to Cepheids (${\rm [Fe/H]}=-0.785$ based on forthcoming work by Romaniello et al., as cited within \citealt{bre24}, and see also \citealt{rom08}).  A similar systematic offset appears amongst IC1613 red and blue supergiant estimates \citep[e.g., compare \citealt{chu22} and][]{bre07}. Despite diffusion potentially altering abundances inferred from B and A-type supergiants, the estimates are favored owing partly to added consistency relative to Cepheids.

The IC1613 distance is linked to the mean between absolute Wesenheit relations tied to and foregoing LKH-like corrections \citep[][and data therein, plus an interpolated value for SY Aur]{ben07,rie18plx}: $W_{VI,0}=(-3.30\pm0.03) \log{P}+(-2.46\pm0.06)$. Half the offset was folded into the zeropoint uncertainty ($0^{m}.04/2$), together with $\pm 0^{m}.05$ linked to $R$.  For the slope term $\pm0.03$ was adopted to encompass Magellanic Cloud Cepheids \citep[Table 4 in][]{bre22}. Certain researchers questioned LKH-like applications in comparable instances, such as M.~Feast \citep[][see passages therein]{smi03,ben07}, \citet{smi03}, \citet{tur10}, and \citet{fra14}. \citet[][]{rie18b} preferred an alternate approach to the broader topic \textit{vis \`a vis} \citet{rie18plx}.  The author's (D.M.) view may appear in a subsequent study (no correction and focusing on separate systematics).

$\gamma (W_{VI})=-0.01 \pm 0.08$ mag dex$^{-1}$ emerges after applying a weighted mean of IC1613 distances (EDD and CCHP), the new IC1613 Cepheid distance, and including $\pm 0.05$ dex uncertainty associated with the metallicity baseline. Importantly, while $\gamma$ could be attached to the Galactic calibration once the uncertainty is decoupled and reduced to avoid redundant contributions, ultimately for distances $W_{VI}$ should be replaced by infrared relations to mitigate the impact of $R$ and other factors \citep[e.g.,][]{maj16}, in tandem with potentially transitioning to non-Wesenheit formulations.

\section{Diverse Opinions ($\gamma$)}
Fig.~1 in \citet{bre25} may provide a disputable impression that a sole datum favors a marginal, null, or insignificant metallicity effect \citep[i.e.,][]{mf25}.  Yet the following research should likewise be considered \textit{in sum} by the reader: \citet{uda01}, \citet{pie04}, \citet{riz07}, \citet{bon08,bon10}, \citet{maj11}, \citet{bha16}, \citet{wie17},\footnote{Consider the \citet{bre22} rebuttal.\label{footnote:bre}} \citet{owe22},$^{\ref{footnote:bre}}$ and \citet[][$-0.07\pm0.21$ mag dex$^{-1}$]{yua22}.\footnote{Cautious qualifiers therein.\label{footnote:yua}}   \citet{bre25} convey that certain metallicity estimates were omitted from Fig.~1 owing to concerns expressed therein and by \citet{bre22} \citep[e.g.,][]{wie17}. 

Moreover, Fig.~1 overlooked larger metallicity effect determinations such as those advocated by \citet[][$-0.52\pm0.19$ mag dex$^{-1}$]{rip25} and \citet[][$-0.80\pm0.21\pm0.06$ mag dex$^{-1}$]{sha11}, and for other broader metallicity dependencies (diverse passbands) that are considerable see also \citet[][]{fau15} and Fig.~14 in \citet{ger11}.  Yet \citet{maj11} countered that the sizable $\gamma (W_{VI})=-0.8$ mag dex$^{-1}$ implied an anomalous SMC distance modulus (e.g., $\mu_0 \neq 18.3$), and was disfavored on that basis (see also Table~\ref{table1}). For general comparison, \citet[][]{gra14} advocated for a ``canonical'' SMC distance of $18.95\pm0.07$, and see also the NED-D compilation \citep[e.g.,][]{ste20} and \citet{gra20}.  The spread in broader metallicity effect determinations and varied individual interpretations are ascertainable by \textit{concurrently} also inspecting \citet[][their Fig.~1]{rom08}, \citet{mar09}, \citet[][their Fig.~14]{ger11}, \citet[][their Table 1]{bre22}, \citet{bon24}, and \citet{rip25a}.  

Furthermore, likewise absent when interpreting the aforementioned Fig.~1 is context concerning the insidious degeneracy between photometric contamination and chemical composition \citep[e.g., \S 5 in][]{mac01,maj10,maj11}.  Those galactocentric trends for remote galaxies can overlap, which compromises results based on that methodology. The key topics are crowding and blending,\footnote{More broadly see \citet{su99} and \citet{moc24}.} and \citet{mac01} identified that the \citet{ken98} $\gamma (W_{VI})$ inferred from the galactocentric spread across M101 was impacted by a non-metallicity source.  \citet[][]{maj11} likewise identified that certain Cepheid data linked to M81 are similarly compromised, by process of elimination. The galactocentric degeneracy between photometric contamination and chemical composition exhibited by NGC4258 (M106) and M33 Cepheids is present in Fig.~4 of \citet[][see also concerns expressed by \citealt{bon08}]{maj09c}, who relied on photometry from \citet{mac06} and \citet{sco09}. \citet{cha12} provided pertinent insight into M33 Cepheid blending in their Fig.~5, and \citet[][]{yua22} noted that prior work on NGC4258 did not apply as enhanced contamination corrections. \citet[][]{yua22} relayed the following relative to the initial \citet{mac06} effort and tied mainly to the inner densest NGC4258 fields, ``\textit{$-0.08\pm0.01$, $-0.14\pm0.01$, and $+0.07\pm0.01$ mag in F555W, F814W, and F555W-F814W, respectively (our corrected values are fainter and bluer).}'' Separately, \citet{maj24b} concluded that the \citet[][]{yua22} and \citet[][]{hof16} NGC4258 photometry were discrepant by $W_{VI,0}\simeq 0^{m}.3$. Contamination is likewise observed in globular clusters \citep[][]{maj12f,maj12e,lee14}, toward the Galactic center sightline \citep[Fig.~2 in \citealt{maj10}, and Fig.~5 in][]{maj16}, IC1613 \citep{maj13,maj14}, RR Lyrae in the Magellanic Clouds \citep{maj18,maj20}, and S$H_0$ES applies considerable decontamination adjustments for extragalactic Cepheids \citep[e.g.,][]{rie11}. Although the reliability of the latter's overall approach is debated \citep[e.g.,][see also \citealt{rb23} and \citealt{ri24} for broader S$H_0$ES positions and rebuttals]{efs20,hoy25}. In sum, the previously mentioned Fig.~1 contains datapoints succumbing to the aforementioned degeneracy, and other issues (e.g., \citealt{riz07} and \citealt{mf23} contested the \citealt{sak04} result). 

Lastly, this current effort underscores that pairing HST parallaxes \citep{ben02,ben07,rie14,rie18plx} can be advantageous while aiming to circumvent debated bright Gaia astrometry. Hipparcos distances to the Pleiades and Blanco 1 are stark reminders to incessantly benchmark emerging satellite data \citep[e.g.,][]{tb02,maj11e}. \citet{bk19} expressed potential reservations regarding the \citet{ben07} HST parallaxes when assessing Gaia DR2, and \citet{bre25} discouraged their use in lieu of DR3. An alternative position is adopted here. Gaia releases thus far continue to exhibit underestimated uncertainties \citep[e.g.,][]{rip19,lin21}, and \citet{bre20} provide a summary of DR2 offsets in their Table~5.  By using DR2 in part \citet{bk19} and \citet[][]{rie18b} arrived at separate $H_0$ determinations ($H_0=69\pm2$ km/s/Mpc by the former), and compare the \citet[][]{bk19} and \citet{bre20} conclusions.  Furthermore, \citet[][]{rie18plx,rie18b} relied partly on the \citet{ben07} FGS HST parallaxes to help inform their distance scale research \citep[see also Fig.~5 in][]{tur10}, in tandem with Hipparcos and their own HST parallaxes \citep{rie14,rie18plx}.  Expectedly, \citet[][]{lin21} implied that seeking such or other (in)validating non-Gaia observations is desirable \citep[e.g.,][and even Gaia data less susceptible to certain biases are pertinent, e.g., putative fainter companions, \citealt{ker19}]{ker14,bob25}.  An important example of a Gaia DR3 discrepancy is relative to the \citet{rie18plx} HST parallax for the critical long-period Cepheid S Vul. Specifically, for S Vul \citet[][DR2]{bj18} obtained $d\simeq3.0$ kpc, whereas \citet[][L21+DR3]{bj21} determined $4.3$ kpc, and the \citet{rie18plx} HST parallax yields $3.1$ kpc (utilized here to provide a Gaia-independent viewpoint).  Other examples possibly include Polaris B \citep{use08} and SV Vul \citep{rie21}. Broader unresolved concerns endemic to the Gaia DR3 release are actively discussed in the literature \citep[e.g., Fig.~10 in][Figs.~6, 7, and 8 in \citealt{kha23}, and also weigh contrasting views by \citealt{owe22} and \citealt{bre22}]{mol23}.

Concluding remarks from \citet{lin21} are worth reiterating, ``\textit{While it is easy enough to demonstrate that the EDR3 parallaxes contain significant systematics ...~This paper does not claim to give a definitive answer but merely a rough characterization ...~better and possibly quite different estimates can be obtained in the future by means of more refined and comprehensive analysis methods. Continued exploration of the systematics is important ...~In the end, this will hopefully lead to much lower levels of systematics in future Gaia data releases.}''   

\section{Conclusion}
Gaia-independent \citet{rie14,rie18plx} and \citet{ben02,ben07} HST parallaxes were paired with IC1613 photometry to benchmark proposed $\gamma (W_{VI})$ (Table~\ref{table1}), since the Gaia zeropoint is debated \citep[][discussion and references therein]{mol23}. Larger corrections are disfavored by a weighted mean of TRGB-JAGB IC1613 distances from \citet[][EDD]{ana21} and \citet[][CCHP]{lee24}. Figs.~2 and 3 in \citet{mf23}, supported by semi-independent corroborating evidence from Table~\ref{table1}, may aid to distinguish which selected model parameters provide realistic outcomes \citep[e.g.,][]{bm99,and16,des22,kha25}. 

At least three outstanding issues remain with the current analysis, and which could motivate subsequent efforts.  A small metallicity effect of $\gamma (W_{VI}) \stackrel{<}{\sim} -0.1$ mag dex$^{-1}$ falls within that deduced here of $\gamma (W_{VI})=-0.01 \pm 0.08$ mag dex$^{-1}$, and more broadly for example:~\citet[][]{maj13e} derived $\gamma (3.6)\simeq-0.10\pm0.10,-0.01\pm0.06$ \citep[see also][$-0.09\pm0.29$ mag dex$^{-1}$]{fre11}; \citet{rie11,rie16} obtained $\gamma (W_{H-VI})=-0.10\pm0.09,-0.14\pm0.06$; \citet{wie17} deduced $\gamma (W_{VI})=-0.025\pm0.067$;$^{\ref{footnote:bre}}$ \citet{gro13} noted $\gamma (W_{VK})=0.04\pm0.10$; and \citet{mf23} established $\gamma=-0.046\pm0.019,-0.022\pm0.015$ mag dex$^{-1}$ $\rm[O/H]$ (sample size dependent and they provide arguments supporting the latter). To semi-independently vet such corrections a desirable aim is achieving the baseline established in Fig.~2 of \citet{mf23} (see also the pertinent argument conveyed in Fig.~20 of \citealt{mf25}). Second, \citet[][]{rie14,rie18plx} and \citet[][]{ben07} parallaxes exhibit certain deviations relative to brighter DR3: e.g., SS CMa, WZ Sgr, SY Aur, FF Aql, RT Aur, S Vul, and Table 1 in \citet[][]{van07} hosts a comparison between Hipparcos and \citet{ben07} HST parallaxes (e.g., Y Sgr) \citep[see also concerns expressed by][]{bk19,bre25}. Ultimately forthcoming DR4 and additional non-Gaia parallaxes may clarify where the issues reside, \textit{once}  DR4 is validated (is DR3 problematic, or instead \citealt{ben07}, or \citealt{rie14,rie18plx}, or combinations thereof). Third, the EDD and CCHP comparison distances for IC1613 may be inaccurate, although their estimates are comparable ($2\times$TRGB, $1\times$JAGB). Type II Cepheids may provide lucrative corroborating or discrepant insight pending further observations \citep[e.g.,][]{maj09c,nar25,das25}.

A disagreement emerged relative to the contention of a converging broad consensus regarding how composition impacts Cepheid distances, where Fig.~1 in \citet{bre25} could propagate a disputed impression that a sole instance exists of a relatively null, marginal, or insignificant metallicity effect, when there is a body of work that \textit{in sum} support that position (\S 3).  Nor was it expressed that photometric contamination critically impacts specific datapoints in that aforementioned figure, and certain research advocating sizable $\gamma (W_{VI})$ was overlooked.  Recent $\gamma (W_{VI})$ estimates by \citet[][$-0.335\pm0.059$ mag dex$^{-1}$]{gie18}, \citet{mf23,mf25}, \citet{bre24}, and \citet{rip25} delineate a $\Delta \gamma (W_{VI})\simeq 0.5$ mag dex$^{-1}$ baseline.  More broadly \citet{rip25a} advocated their interpretation that there is no consensus regarding $\gamma$, which matches the position espoused here.  

\begin{acknowledgments}
This research relied on initiatives such as CDS, NASA ADS, arXiv, OGLE, Gaia, (C)CHP, S$H_0$ES, EDD, Araucaria, Gaia, C$-$MetaLL, Hipparcos. 
\end{acknowledgments}

\bibliography{article}{}
\bibliographystyle{aasjournalv7}

\end{document}